\documentclass[reprint,aps,prl]{revtex4-1}
\pdfoutput=1

\usepackage{amsmath}
\usepackage{physics}
\usepackage{color}
\usepackage{microtype}
\usepackage{subcaption}
\usepackage{graphicx}

\newcommand\comment[1]{} 

\usepackage{hyperref}
\hypersetup{
    bookmarksnumbered=true, 
    unicode=false, 
    pdfstartview={FitH}, 
    pdftitle={}, 
    pdfauthor={}, 
    pdfsubject={}, 
    pdfcreator={}, 
    pdfproducer={}, 
    pdfkeywords={}, 
    pdfnewwindow=true, 
    colorlinks=true, 
    linkcolor=blue, 
    citecolor=blue, 
    filecolor=blue, 
    urlcolor=blue 
}

\usepackage{algorithmicx}
\usepackage{amsthm}
\usepackage{amsfonts,amssymb}

\newtheorem{theorem}{Theorem}

\begin{document}
\date{\today}
\title{Parallel Position-Controlled Composite Quantum Logic Gates with Trapped Ions}
\author{Michael S. Gutierrez}
\email[email:]{m\_g@mit.edu}

\author{Guang Hao Low}
\author{Richard Rines}
\author{Helena Zhang}
\affiliation{Center for Ultracold Atoms, Department of Physics, Massachusetts Institute of Technology, Cambridge, Massachusetts 02139, USA}

\begin{abstract}
We demonstrate parallel composite quantum logic gates with phases implemented locally through nanoscale movement of ions within a global laser beam of fixed pulse duration. We show that a simple four-pulse sequence suffices for constructing ideal arbitrary single-qubit rotations in the presence of large intensity inhomogeneities across the ion trap due to laser beam-pointing or beam-focusing. Using such sequences, we perform parallel arbitrary rotations on ions in two trapping zones separated by 700~$\mu$m with fidelities comparable to those of our standard laser-controlled gates. Our scheme improves on current transport or zone-dependent quantum gates to include phase modulation with local control of the ion's confinement potential. This enables a scalable implementation of an arbitrary number of parallel operations on densely packed qubits with a single laser modulator and beam path. 
\end{abstract}

\maketitle

Quantum processors with hundreds or more qubits promise to deliver significant computational speedups over the best classical systems \cite{VanMeter2013}. The limits of physical coherence make large-scale parallelization of primitive quantum operations crucial for realizing a fault-tolerant device~\cite{Gottesman2014}. Trapped ions are one of the leading candidates for physical qubits due to their consistency---all ions are identical---and large ratio of coherence \cite{Ruster2016} to gate \cite{Mizrahi2014} times. Few-qubit ion systems can now demonstrate simple quantum algorithms~\cite{Monz2016, Debnath2016} as well as single- and multi-qubit operations within the fault-tolerant regime~\cite{Ballance2016, Harty2016,Linke2016,Blume-Kohout2016}.

    However, current ion trap systems rely upon bulky free-space optical components and high-power radio-frequency laser modulators, both of which pose daunting technical difficulties~\cite{Haffner2008,Frohlich2007} to scaling to hundred- or thousand-qubit parallel systems \cite{Meter2006}. A major challenge moving forward is managing and optimizing physical resources required to implement high-performing quantum operations at scale. To this end, many proposed architectures identify key resources that offer clear and ready paths toward scaling up to an arbitrary number of qubits \cite{Brown2016, Harlander2012,Leibfried2007}.
    
   	One promising resource is fine voltage control of trap electrodes, which can be harnessed to displace the confining potential of single ions. Local targeted qubit operations have been performed using potential displacement in conjunction with static laser interaction zones~\cite{Harlander2012, Leibfried2007,Clercq2016,Nigmatullin2016} or magnetic field gradients~\cite{Warring2013, Weidt2016}. These schemes can greatly reduce complexity of optical addressing systems, and replace the numerous high-power laser modulators with low-power voltage generators which can be readily integrated on-chip with existing technology~\cite{Mehta2014}. However, quantum control techniques proposed thus far have focused on using local voltage changes to gate the interaction time by transporting ions to or through designated operation zones. This approach requires each ion to be transported over large distances ($>$100~$\mu$m), greatly limiting the speed and density of parallel operations.

	In this Letter, we propose an alternative approach using nanoscale ion movements parallel to the laser beam to implement local phase changes of the global beam. This scheme offers several significant advantages over previous works. First, the number of parallel ion movements per beam pass is limited only by the number of independently controlled electrodes, which is highly scalable. Second, movement operations are local and space efficient: ions remain within a single trapping zone and only undergo sub-micron displacements. Third, we demonstrate position-controlled composite sequences that enable arbitrary and ideal single-qubit operations on each ion in parallel, despite the large inhomogeneities that can arise in a global beam. 

	We implement our scheme experimentally using easily scalable trapping and control technologies. We first apply local, position-controlled phases to a single ion using a fixed-frequency, fixed-phase laser beam to perform a Ramsey experiment. To enable parallel operations despite zone-dependent global beam intensity, we construct a simple and efficient four-pulse composite sequence to generate ideal arbitrary single-qubit rotations. We prove the scalability of such an approach by performing parallel single-qubit operations on two ions in independent trapping zones separated by 700~$\mu$m with no additional optics, modulators, or timing overhead. We find the fidelities of these parallel gates to be comparable to standard, optically modulated gates. We describe the limitations for these nanoscale movements in our current setup and find them comparable in timing and phase precision to typical direct digital synthesizers (DDSs) used for laser modulation.

\begin{figure}[ht]
	\includegraphics[width=\columnwidth]{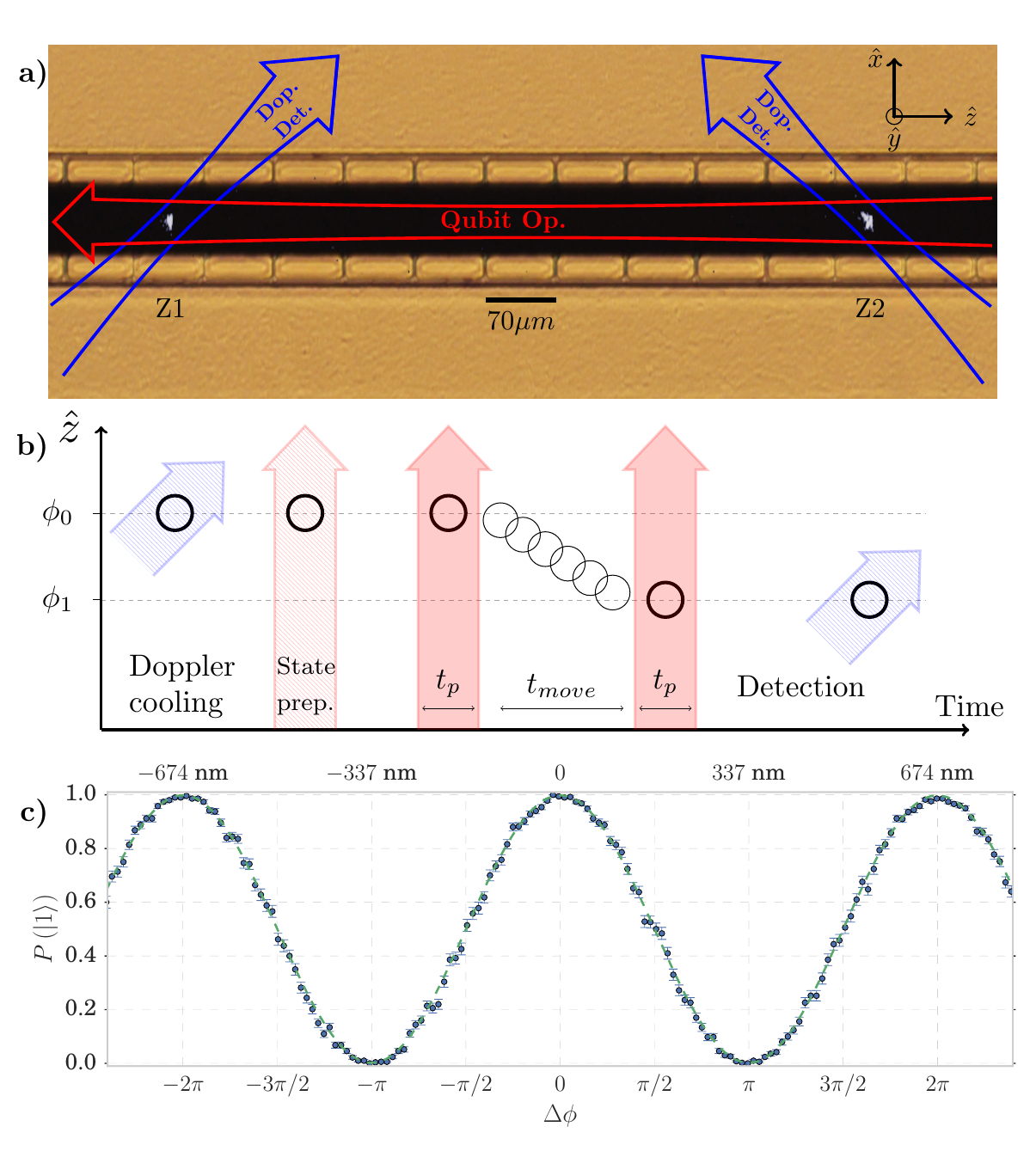}
	\caption{(a) Image of the surface trap indicating geometries of Doppler, detection, and qubit operation beams. The two trapping zones of interest, Z1 and Z2, are labeled with ion fluorescence shown. (b) The Ramsey timing sequence, with pulse duration $t_p$ = 1.5~$\mu$s and movement duration $t_{move}$ = 8~$\mu$s for a single ion (black circle) in one zone. Blue arrows represent zone-addressed Doppler cooling and state-detection. Red arrows indicate global qubit operations (solid) and global state preparation (shaded). (c) Experimental Ramsey sequence data, showing the $\ket{1}$ population vs. distance moved from start position expressed in phase, with 500 repeats per data point. The fit gives a $99.6 \pm 0.3 \%$ contrast and a voltage-distance calibration within $1\%$ of boundary-element-method simulation results. }
	\label{fig:exp}%
\end{figure}

	Our qubit of choice is stored in the electronic states $\ket{ 0 } = 5S_{1/2}[m=-1/2]$ and $\ket{1} = 4D_{5/2}[m=-5/2]$ of the Strontium ion. We confine single ions 68~$\mu$m above a Sandia Laboratories High Optical Access trap \cite{Maunz2016}. A 4~G$\hat y$ static uniform magnetic field is used to spectrally isolate these Zeeman states, where the $\hat y$-axis is normal to the trap surface. Qubit states are manipulated using a narrow linewidth $\lambda =$ 674~nm laser passing along the trap axis of symmetry ($\hat z$-axis) with a waist of $w_0 =$~25~$\mu$m. The $\hat z$-secular frequency is $2\pi \times 1.25$~MHz. The confining potential is generated via 188 control surfaces connected to a 48-channel home-built arbitrary waveform generator (AWG) with $\pm$10~V output range, 20-bit precision, and 250 kHz asynchronous update rate, enabling sub-nanometer position control. Voltages from the AWG are low-pass filtered at 600 kHz en route to the trap electrodes. Two trapping zones separated by 700~$\mu$m, Zone 1 (Z1) and Zone 2 (Z2), are used throughout this work and depicted in Fig. \ref{fig:exp}a along with beam geometries.
    
	Each experimental sequence is preceded by 300~$\mu$s of zone-addressed Doppler cooling,  100~$\mu$s of global frequency-selective optical pumping and 500~$\mu$s of sideband cooling \cite{Schindler2013}, preparing the ion in the $\ket{0}$ state with $\sim$$99.5\%$ fidelity. After the sequence, zone-addressed state detection is performed by recording state-dependent fluorescence with a single-photon photomultiplier tube \cite{Myerson2008}, elapsing 200 $\mu$s and with readout fidelity of $\sim$$99.9\%$. The amplitude of qubit operations are calibrated by measuring Rabi oscillations at Z1 and Z2 to the $\sim$$1\%$ level, yielding zone-dependent Rabi frequencies of $\Omega_{Z1} = 2\pi\times166$~kHz and $\Omega_{Z2} = 2\pi\times159$~kHz.

	Local phase control of qubit operations is achieved by moving each ion along the $\hat z$-axis between identical global laser pulses of length $t_p=1.5$~$\mu$s, as depicted in Fig.~\ref{fig:exp}b.  We demonstrate phase control by performing a Ramsey experiment on a single ion, varying the displacement of the confining potential, with results shown in Fig.~\ref{fig:exp}c. The displacement is generated with a single voltage set update, with an 8~$\mu$s delay between $\pi/2$ pulses to allow for voltages to settle before subsequent pulses. The resulting $\sin^2(\Delta\phi)$ behavior calibrates voltages to the ion's displacement $\Delta z$ via the observed phase shift $\phi= 2\pi \Delta z / \lambda$, which is within $\sim$1$\%$ of boundary element method simulations \cite{Maunz2016}. 

	The Ramsey sequence suffices for accessing the full $4\pi$ rotation space in a single zone. However, fully parallel operations with $\mathcal{O}(10^{-4})$ infidelity would require intensity differences of no more than $\sim$1$\%$ across the trap. While it is possible to reduce alignment error sufficiently, this would still decrease the usable interaction region of a Gaussian beam to one-hundredth of a Rayleigh range. For our beam geometry, this would reduce the usable interaction region to just 30~$\mu$m, less than $1\%$ of the 4~mm trapping region of our device.
    
\begin{figure}[t!]
	\includegraphics[width=\columnwidth]{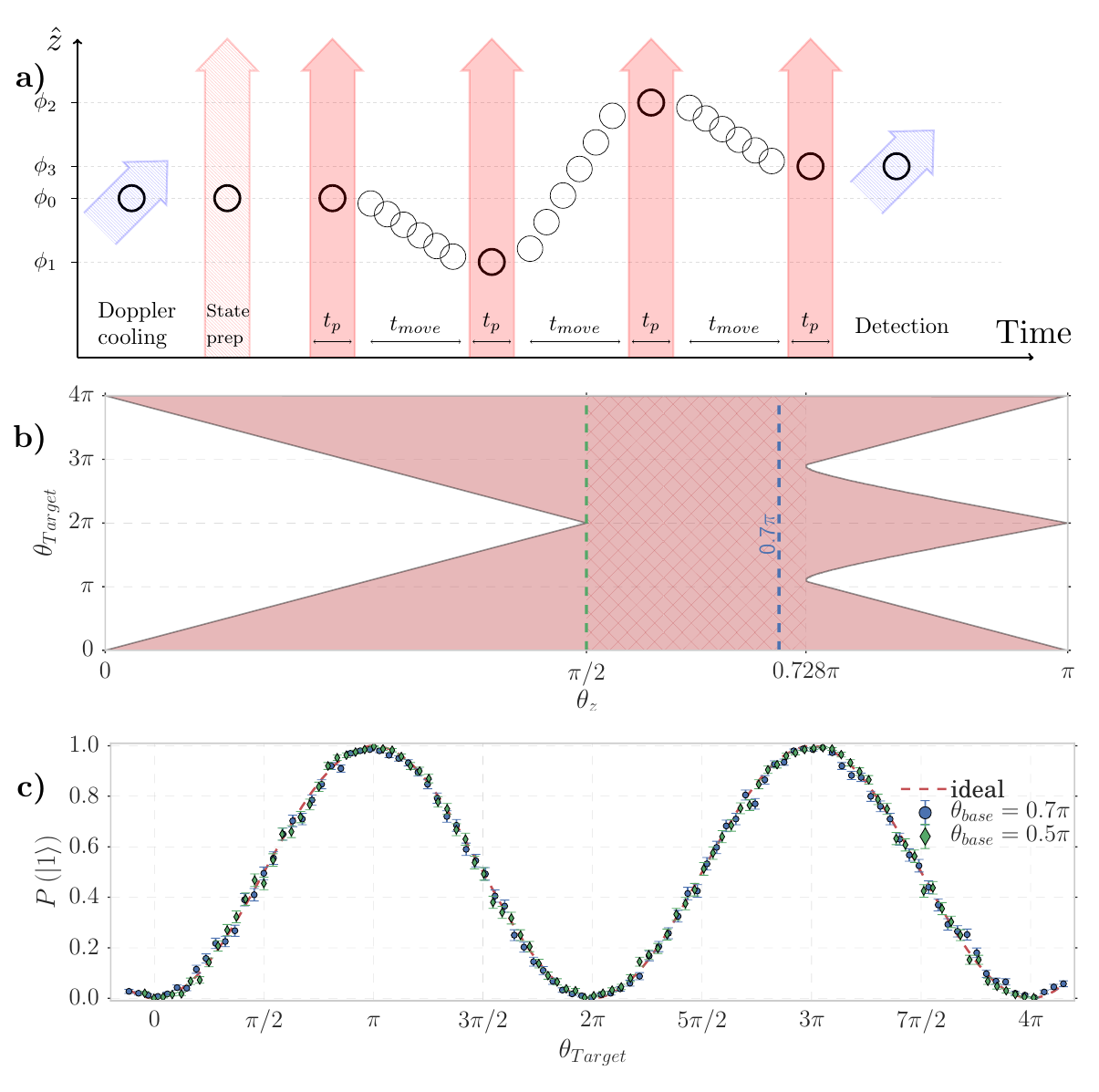}
	\caption{(a) Composite four-pulse sequence and movement timing, with each experiment pulse lasting $t_p = 1.5$~$\mu$s and each movement period lasting $t_{move} = 8$~$\mu$s for a total gate time of 30~$\mu$s. (b) Composite sequence solutions space: shaded red region indicates a four-pulse solution exists for the desired target rotation ($\theta_{\text{Target}}$) and base rotation ($\theta_z$), and cross-hatched region is where the full $4\pi$ target rotation is achievable. Angle $\theta_z=0.7\pi$, corresponding to approximately twice the intensity at $\theta_z=\pi/2$, is indicated. (c) Experimental data for the $\ket{1}$-state population vs target rotation for two base rotations $\theta_z = \pi/2$ (circles) and $\theta_z = 0.7 \pi$ (diamonds), with contrast given by $99.1 \pm 0.5 \%$ and $98.0 \pm 0.5 \%$ respectively.
	}
	\label{fig:pulse}
\end{figure}

    To extend this range, we construct a composite quantum gate of fixed pulse durations and variable phases following the methodology of Low et.~al.~\cite{Low2016}. We require that all possible ideal single-qubit rotations are achievable by the composite gate within a continuous range of Rabi frequencies such that we are completely insensitive to beam intensity inhomogeneities across the trap. The shortest composite gate satisfying this constraint is a sequence of length four. The total duration of the laser pulses is $2\pi$, so the sequence does not add excessive overhead. Details for extracting phases for each of the four pulses and generating a specific target rotation $\theta_{\text{Target}}$ given a base rotation $\theta_z = \int_0^{t_p}\Omega_z dt$ are outlined in the Supplemental Material. We define the region-of-validity to be the range of base and target rotations in which the constraint is satisfied, shown for the length-four sequence in Fig.~\ref{fig:pulse}b. Observe that the full $4\pi$-rotation space is achievable for base rotations varying between $\theta_z \in \{ \pi/2, 0.728\pi \}$ or equivalently, intensity variations between $\{I_{min}, 2.12I_{min} \}$. This extends the usable interaction region to two Rayleigh ranges, two hundred times larger than that of the basic Ramsey sequence. With our beam geometry, this now covers the entire length of the trapping region.

	To empirically demonstrate the region-of-validity of the four-pulse composite gate, we implement scans of the target rotation from 0 to $4\pi$ using two different base rotations: one along the minimum boundary,  $\theta_z = \pi/2$, and one at approximately twice that intensity, $\theta_z = 0.7 \pi$, as shown in Fig.~\ref{fig:pulse}c. Timing for each sequence is identical and is shown in Fig.~\ref{fig:pulse}a. The contrast of each scan is $99.1\pm 0.5\%$ and $98.0 \pm 0.5\%$, respectively, consistent with errors in our typical laser-controlled operations.

\begin{figure}[t!]
\includegraphics[width=\columnwidth]{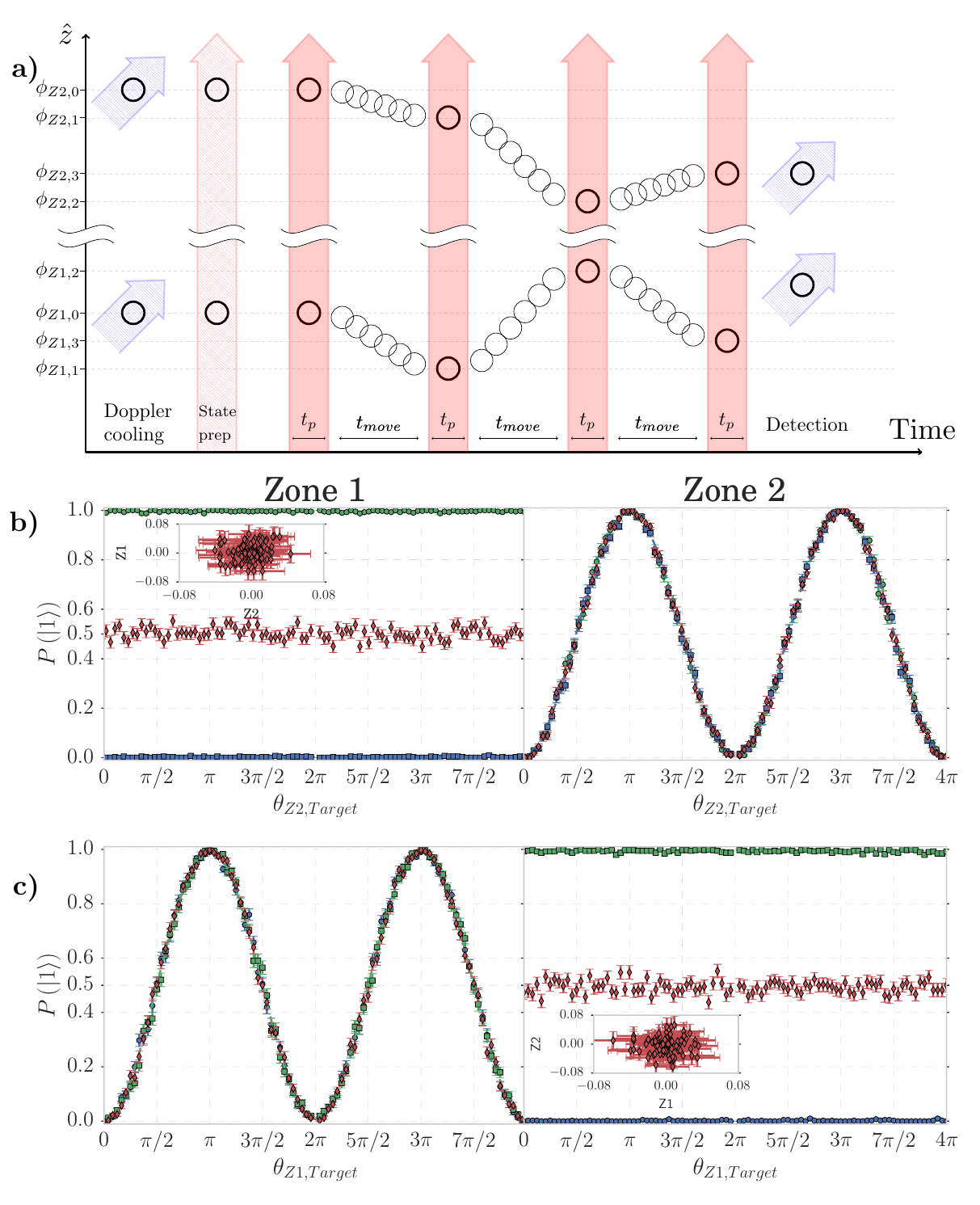}
\caption{ (a) Two-zone parallel composite sequence pulse and movement timing. Doppler cooling, state-prep, experimental sequence and state detection are all performed simultaneously, with $t_p=1.5$~$\mu$s and $t_{move}=8$~$\mu$s. (b) Experimental data for Z1 and Z2 $\ket{1}$-population for a Z2 target rotation scan and a constant Z1 target rotation: Identity (diamonds), $\pi/2$ (circles) and $\pi$ (squares) (c) Experimental data for Z1 and Z2 $\ket{1}$-population for a Z1 target rotation scan and a constant Z2 target rotation: Identity (diamonds), $\pi/2$ (circles) and $\pi$ (squares). (b-c) insets show Z1 and Z2 residuals for the $\pi/2$-target rotation.}
\label{fig:TwoZones}
\end{figure}

	The wide range of intensities at which the composite gates can be performed offers the ability to parallelize many operations with a single beam pass. We show the ability to perform independent parallel composite operations in two zones separated by 700~$\mu$m. In order to minimize cross-talk, the confinement potential is solved to satisfy simultaneous constraints for both zones while specifying position, $\hat z$-secular frequency, tilt in $xy$-plane, and stray-field compensation using a fast multipole expansion for each collection of AWG-controlled electrodes~\cite{Singer2010}, accounting for fields from both neighboring electrodes which share the same AWG and electrodes from the other trapping zone. We then scan the target rotation in one zone while implementing one of three target rotations in the other: identity, $\pi/2$ rotation, or $\pi$ rotation, with the results shown in Fig.~\ref{fig:TwoZones}. For each case, the scanned target rotation shows a contrast of $99.2 \pm 0.4 \%$ and $99.1 \pm 0.5\%$ for Z1 and Z2 respectively. The constant target rotation is within statistical uncertainty of the desired operation for each case. We further look for cross-talk between the gates by computing the correlation between the residuals of Z1 and Z2 data for each constant target rotation compared with the ideal target rotation. We find that all normalized covariances are insignificant, with $\norm{ \frac{ \mathrm{Cov}(Z1,Z2) }{ \sqrt{ \mathrm{Var}(Z1) \mathrm{Var}(Z2) } } } < 0.08$. To further characterize the fidelity of each gate and the cross talk would require randomized benchmarking \cite{Knill2008} or gate set tomography \cite{Blume-Kohout2013}, which is beyond the scope of this work.
    
    The fundamental limits to the performance of these gates will be set by the phase resolution achievable for the specific trap geometry and AWG used. To estimate the phase resolution achievable in our current setup, we simulate the electric fields generated by the ion's nearest-neighbor electrodes to be 0.25~V/mm. Combined with our AWG resolution and $\hat z$-secular frequency, this results in roughly 12 bits of resolution over one $\lambda$ of movement or approximately $0.0015$~rad over a $2\pi$ phase shift. Further resolution could be achieved through increased $\hat z$-secular frequency, optimized trap designs, or increased AWG resolution. However, we note that the resolution achieved here is already comparable to most DDSs used for laser modulation, which have 10 to 16 bits of phase precision.
    
    Limitations of a more technical nature will occur due to the motional excitation induced from the cascade of ion movements. In this work, the relatively slow 8~$\mu$s movement time keeps the motion in the adiabatic regime. However, several groups have already demonstrated ion transport that is both fast and induces minimal motional excitation \cite{Walther2012,Fadel2016,Bowler2012}. Incorporating these works and increasing AWG update rates would reduce the movement time to sub-microsecond levels, comparable to typical DDS programming times.
    
    Beyond technical improvements, the composite gate itself can be extended by looking at longer sequence lengths, adding constraint equations for more exotic quantum response functions, or extending to non-equiangular sequences. Longer sequences will offer an increased range-of-validity, allowing more tightly focused beams or longer traps. Incorporating additional constraints can allow for error-resilient broadband rotations and compensate for architecture-specific errors, such as detuning and imperfect movement. Non-equiangular composite sequences can compensate for intensity gradients over the $\lambda$-movement range, making this work applicable to tightly-focused radial beams and thus adding phase control to prior work such as de Clercq et.~al.~\cite{Clercq2016}. Combined with already demonstrated work incorporating CMOS compatibility~\cite{Mehta2014} and integrated optics~\cite{Mehta2015}, this scheme would enable a completely integrated trapped ion system, avoiding the need for complex, active optics. 

	In summary, we have demonstrated parallel position-controlled composite quantum logic gates, where phases are implemented by nanoscale movements of each ion within a global laser beam. We described how to overcome imperfect rotations caused by zone-dependent light intensities through the construction of a four-pulse composite gate, allowing the intensity to vary by more than a factor of two between interaction zones. Our scheme provides a pathway toward dense parallelized quantum operations on ions with minimal optics and external modulators.

	We would like to thank Peter Maunz and Daniel Stick for useful discussions and providing documentation for the High Optical Access trap. This work was funded in part by the IARPA MQCO program and by the NSF Center for Ultracold Atoms. Guang Hao Low would like to acknowledge funding by the ARO Quantum Algorithms Program and NSF RQCC Project No.1111337.

\bibliographystyle{apsrev4-1}

\bibliography{bib/MotionGate}

\appendix
\begin{widetext}
\section{Supplemental Material: Composite Quantum Gates}

A principled and systematic approach to the design of composite quantum gates is surprisingly challenging. Given a sequence of $L$ primitive rotations $\hat{R}_\phi[\theta]$, the essential problem is finding an intuitive characterization of the functional form of all possible composite quantum gates $\hat{U}[\theta]$ of the form 
\begin{align}
\label{Eq:UnitarySequence}
\hat{U}[\theta]=\hat{R}_{\phi_{L-1}}[\theta]\cdots\hat{R}_{\phi_1}[\theta]\hat{R}_{\phi_0}[\theta],\quad \hat{R}_{\phi_0}[\theta]=e^{-i\frac{\theta}{2}(\cos{(\phi_0)}\hat{\sigma}_x+\sin{(\phi_0)}\hat{\sigma}_y)}.
\end{align}
Note that $\hat{U}[\theta]$ is an $\text{SU}(2)$ operator, thus it can always be decomposed into the Pauli basis $\{\hat{1},\hat{\sigma}_x,\hat{\sigma}_y,\hat{\sigma}_z\}$:
\begin{align}
\hat{U}[\theta]=A[\theta]\hat{1}+i(B[\theta]\hat{\sigma}_z+C[\theta]\hat{\sigma}_x+D[\theta]\hat{\sigma}_y),
\end{align}
where $A[\theta],B[\theta],C[\theta],D[\theta]$ are real functions of $\theta$. Though one approach to this problem is to find the best-fit $\vec{\phi}\in\mathbb{R}^L$ to some objective for the $A,B,C,D$ by gradient-descent, this approach is not particularly insightful, and has an exponentially increasing computational complexity with respect to $L$.

An alternative approach arises from noting that the fidelity $\mathcal{F}^2$ of $\hat{U}[\theta]$ with respect to a target single-qubit rotation $\hat{R}_0[\theta_{T}]$ depends only on the functions $A[\theta]$ and $C[\theta]$. 
\begin{align}
\label{Eq:Fidelity}
\mathcal{F}=\frac{1}{2}\left|\tr{\hat{R}_0[\theta_{T}]\hat{U}^\dag[\theta]}\right|=\left|\cos{\left(\frac{\theta_T}{2}\right)}A[\theta]-\sin{\left(\frac{\theta_T}{2}\right)} C[\theta]\right|.
\end{align}
Thus designing $\hat{U}[\theta]$ to implement $\hat{R}_0[\theta_{T}]$ at a specific value $\theta=\theta_0$ or in its neighborhood  requires an understanding of what functions $A[\theta],C[\theta]$ can be implemented by some choice of $\vec{\phi}\in\mathbb{R}^L$. Note that the $\hat{\sigma}_z$ rotations required to implement $\hat{R}_\phi[\theta_{T}]$ are trivially obtained by a global shift of all $\phi_k\rightarrow \phi_k + \phi$. This understanding was previously provided by one of the authors~\cite{Low2016}: 
\begin{theorem}
\label{Thm:Characterization}
A choice of $A[\theta],C[\theta]$ is achievable by some $\vec{\phi}\in\mathbb{R}^L$ if and only if all the following are true\\
(1) $A[0]=1$, \\
(2) $A^2[\theta]+C^2[\theta]\le 1$, \\
(3; $L$ odd) $A[\theta]=\sum^{L}_{k\;\text{odd}}a_k \cos^k{(\theta/2)},\quad \forall k,\; a_k\in\mathbb{R}$, \\
(4; $L$ odd) $C[\theta]=\sum^{L}_{k\;\text{odd}}c_k \sin^k{(\theta/2)},\quad \forall k,\; c_k\in\mathbb{R}$.\\
(3; $L$ even) $A[\theta]=\sum^{L}_{k\;\text{even}}a_k \cos^k{(\theta/2)},\quad \forall k,\; a_k\in\mathbb{R}$, \\
(4; $L$ even) $C[\theta]=\cos{(\theta/2)}\sum^{L}_{k\;\text{odd}}c_k \sin^k{(\theta/2)},\quad \forall k,\; c_k\in\mathbb{R}$.\\
Moreover, $\vec{\phi}\in\mathbb{R}^L$ can be computed from $A[\theta],C[\theta]$ in $\text{poly}(L)$ time.
\end{theorem}
These constraints are particularly intuitive. (1) arises from considering $\theta=0$. There, $\hat{U}[0]$ is identity, thus $A[0]=1$. (2) is simply a statement that probabilities are bounded by $1$. (3,4) arise from a direct expansion of Eq.~\ref{Eq:UnitarySequence} and restrict the form of $A[\theta],C[\theta]$ to simple functions. 

In other words, $A[\theta],C[\theta]$ are trigonometric polynomials in $\cos{(\theta/2)}$ and $\sin{(\theta/2)}$ with a bounded norm. Using the Chebyshev polynomials of the first and second kind $T_k[\cos{(\theta)}]=\cos{(k\theta)}$ and $U_k[\cos{(\theta)}]=\frac{\sin{((k+1)\theta)}}{\sin{(\theta)}}$, the case of even $L$ can be simplified to 
\begin{align}
A[\theta]&=\sum^{L}_{k\;\text{even}}a_k \cos^k{(\theta/2)} =\sum^{L/2}_{k=0}a'_k \cos{(k\theta)}, 
\quad C[\theta]=\cos{(\theta/2)}\sum^{L}_{k\;\text{odd}}c_k \sin^k{(\theta/2)}=\sum^{L/2}_{k=1}c'_k \sin{(k\theta)}.
\end{align}
Thus $A[\theta],C[\theta]$ are any Fourier series with a bounded norm. In the following, all even $L$ sequences are represented by this Fourier series, and so we drop the primes on the coefficients.

As the implementing phases $\vec{\phi}$ can be efficiently computed from $A[\theta],C[\theta]$, it suffices to specify the composite quantum gate only through some choice of $A[\theta]$ and $C[\theta]$. This representation has the significant advantage of very directly describing the fidelity response function of the composite quantum gate, in contrast to $\vec{\phi}$, which provide no direct information about what the implemented composite gate does. It is also often the case that the expression for the trigonometric polynomials or Fourier series $A[\theta],C[\theta]$, as a function of say $(\theta_0,\theta_T)$, is much simpler than that of $\vec{\phi}$. Note however that multiple different, but valid, solutions of $\vec{\phi}$ can be obtained for each choice of $A[\theta],C[\theta]$. We now apply this characterization to the design of composite quantum gates.

Our movement-controlled composite gates should ideally satisfy the following properties: 
\begin{enumerate}
\item For some fixed value of $\theta=\theta_0$, there exist $\vec{\phi}$ such that $\hat{U}[\theta_0]=\hat{R}_0[\theta_T]$ for all $\theta_T\in[0,4\pi)$. This ensures that we can implement all possible single qubit rotations in a single sequence. 
\item Property (1) holds for a continuous range of $\theta_0\in[\theta_{\text{min}},\theta_{\text{max}}]$, $0<\theta_{\text{min}}<\theta_{\text{max}}$. This ensures that for all variations in $\theta$ induced by say, an inhomogeneous laser beam, an ideal arbitrary single-qubit gate can still be implemented in a single sequence, so long as $\theta_0$ is known. 
\end{enumerate}

Combined with Thm.~\ref{Thm:Characterization}, the design of our desired composite quantum gates reduces to finding Fourier series $A[\theta],C[\theta]$ that satisfies these properties, which can be contrasted to the more direct, but less efficient and less insightful numerical search for $\vec{\phi}$ for every pair $(\theta_0,\theta_T)$. In particular, Thm.~\ref{Thm:Characterization} and these properties furnish a set of linear constraints on the coefficients $\{a_k,c_k\}$:
\begin{align}
\label{Eq.LinearSystem}
A[0]&=1,&\text{Thm.~\ref{Thm:Characterization}.1}, 
\\ \nonumber
A[\theta_0]&=\cos{(\theta_T/2)}, &\text{Eq.~\ref{Eq:Fidelity}: }\mathcal{F}=1,
\\ \nonumber
C[\theta_0]&=-\sin{(\theta_T/2)}, &\text{Eq.~\ref{Eq:Fidelity}: }\mathcal{F}=1,
\\ \nonumber
\left[\cos{(\theta_T/2)}\frac{dA[\theta]}{d\theta}-
\sin{(\theta_T/2)}\frac{dC[\theta]}{d\theta}\right]_{\theta=\theta_0}&=0,&\text{Eq.~\ref{Eq:Fidelity}: }\left.\frac{d\mathcal{F}}{d\theta}\right|_{\theta=\theta_0}=0.
\end{align}
Thus we have $4$ linear equations for $L+1$ coefficients of terms in $A[\theta],C[\theta]$. All that remains is the choose $L-3$ additional linear equations, say $c_2=0$ that leads to the satisfaction of condition Thm.~\ref{Thm:Characterization}(2). As any fully-determined system of linear equations can be easily solved, we consider a closed-form specification of such a system of linear equations to be equivalent to finding the $A[\theta],C[\theta]$ in closed-form, which is then equivalent, through  Thm.~\ref{Thm:Characterization}, to finding $\vec{\phi}$ in closed-form. Note that many such choices of these remaining linear equation are possible, and could be constructed to impose additional desirable properties such as flatness of the fidelity response function $\mathcal{F}$ with respect to variations in $\theta$.

\section{Length $3$ Composite Gates}
Given the $4$ linear constraints of Eq.~\ref{Eq.LinearSystem}, the shortest composite gate that could possibly satisfy them all simultaneously must have $4$ total coefficients in the $A[\theta],C[\theta]$ terms. This corresponds to $L=3$. When expanded fully, these constraints are
\begin{align}
\label{Eq.LinearSystem3}
1&=a_1+a_3,
\\ \nonumber
0&=a_1 \cos{(\theta_0/2)}+a_3\cos^3{(\theta_0/2)}-\cos{(\theta_T/2)},
\\ \nonumber
0&=c_1 \sin{(\theta_0/2)}+c_3\sin^3{(\theta_0/2)}+\sin{(\theta_T/2)},
\\ \nonumber
0&=\cos{\left(\frac{\theta_T}{2}\right)}\sin{(\theta_0/2)}\left(a_1+3 a_3\cos^2{(\theta_0/2)}\right)+\sin{\left(\frac{\theta_T}{2}\right)}\cos{(\theta_0/2)}\left(c_1+3 c_2\sin^2{(\theta_0/2)}\right).
\end{align}
As these are a system of linear equations, they can solved easily for the coefficients $(a_1,a_3,c_1,c_3)$ which then furnish $A[\theta],C[\theta]$. Provided that $A[\theta],C[\theta]$ satisfy the conditions of Thm.~\ref{Thm:Characterization}, we are guaranteed that the phases $\vec{\phi}$ implementing this gate can be efficiently computed. 

It then remains to determine the parameter space of $(\theta_0,\theta_T)$ such that $A[\theta],C[\theta]$ is achievable. By construction, conditions (1,3,4) of Thm.~\ref{Thm:Characterization} are satisfied by the $A[\theta],C[\theta]$ obtained from Eq.~\ref{Eq.LinearSystem3}. Thus all that remains to guarantee that $\vec{\phi}$ exists is to check condition (2) that $A^2[\theta]+C^2[\theta]\le 1$. We now derive necessary and sufficient conditions for $(\theta_0,\theta_T)$ that satisfy this condition. Let us expand $A^2[\theta]+C^2[\theta]$ about $\theta=0,\theta_0$:
\begin{align}
A^2[\theta]+C^2[\theta]&=1+\frac{\theta^2}{4}\frac{
\left(\cos{(\theta_0/2)}-\cos{(\theta_T/2)}\right)^3\left(\cos{(3\theta_0/2)}-\cos{(\theta_T/2)}\right)}{\sin^2{(\theta_0)}\cos^2{(\theta_0/2)}\sin^2{(\theta_T/2)}}
+\mathcal{O}(\theta^4)\le 1 \\ \nonumber
A^2[\theta]+C^2[\theta]&=1+(\theta-\theta_0)^2\frac{
\left(\cos{(\theta_0/2)}-\cos{(\theta_T/2)}\right)^3\left(\cos{(3\theta_0/2)}-\cos{(\theta_T/2)}\right)}{\sin^2{(\theta_0)}\sin^2{(\theta_T/2)}}
+\mathcal{O}((\theta-\theta_0)^4)\le 1.
\end{align}
By construction through Eq.~\ref{Eq.LinearSystem3}, $A^2[\theta]+C^2[\theta]=1$ at $\theta=0,\theta_0,2\pi-\theta_0$ -- the point $\theta=\pi-\theta_0$ arises from the symmetry $A^2[\theta]+C^2[\theta]=A^2[-\theta]+C^2[-\theta]$ -- and these are also stationary points. By choosing 
\begin{align}
g_3(\theta_0,\theta_T)=\left(\cos{(\theta_0/2)}-\cos{(\theta_T/2)}\right)\left(\cos{(3\theta_0/2)}-\cos{(\theta_T/2)}\right) < 0,
\end{align}
so that the second derivative of $A^2[\theta]+C^2[\theta]$ about $\theta=0$ is negative,
the intermediate value theorem tells us that at least $3$ additional stationary points also develop at $A^2[\theta]+C^2[\theta]<1$ between these values of $\theta$. Thus we have identified at least $6$ stationary points of $A^2[\theta]+C^2[\theta]$ that all have value  $\le 1$. However, as $A^2[\theta]+C^2[\theta]$ is a Fourier series of degree of degree $3$, it has at most $6$ stationary points in $\theta\in[0,2\pi)$. Thus we conclude that $\forall\theta\in \mathbb{R},\;A^2[\theta]+C^2[\theta]\le 1$ if $\{(\theta_0,\theta_T)|g_3(\theta_0,\theta_T) < 0\}$. This can be strengthened to if and only if $\{(\theta_0,\theta_T)|g_3(\theta_0,\theta_T) \le 0\}$, plotted in Fig.~\ref{Fig:SymmetricSequence}, by noting that $A^2[\theta]+C^2[\theta]=1$ when $g_3(\theta_0,\theta_T)=0$.

Observe from Fig.~\ref{Fig:Sequence3} that the full range of $\theta_T\in[-2\pi,2\pi]$ is not achievable for any fixed $\theta_0$. Thus we must explore longer sequences of $\vec{\phi}$.
\begin{figure}[h]
\includegraphics[width=0.5\textwidth]{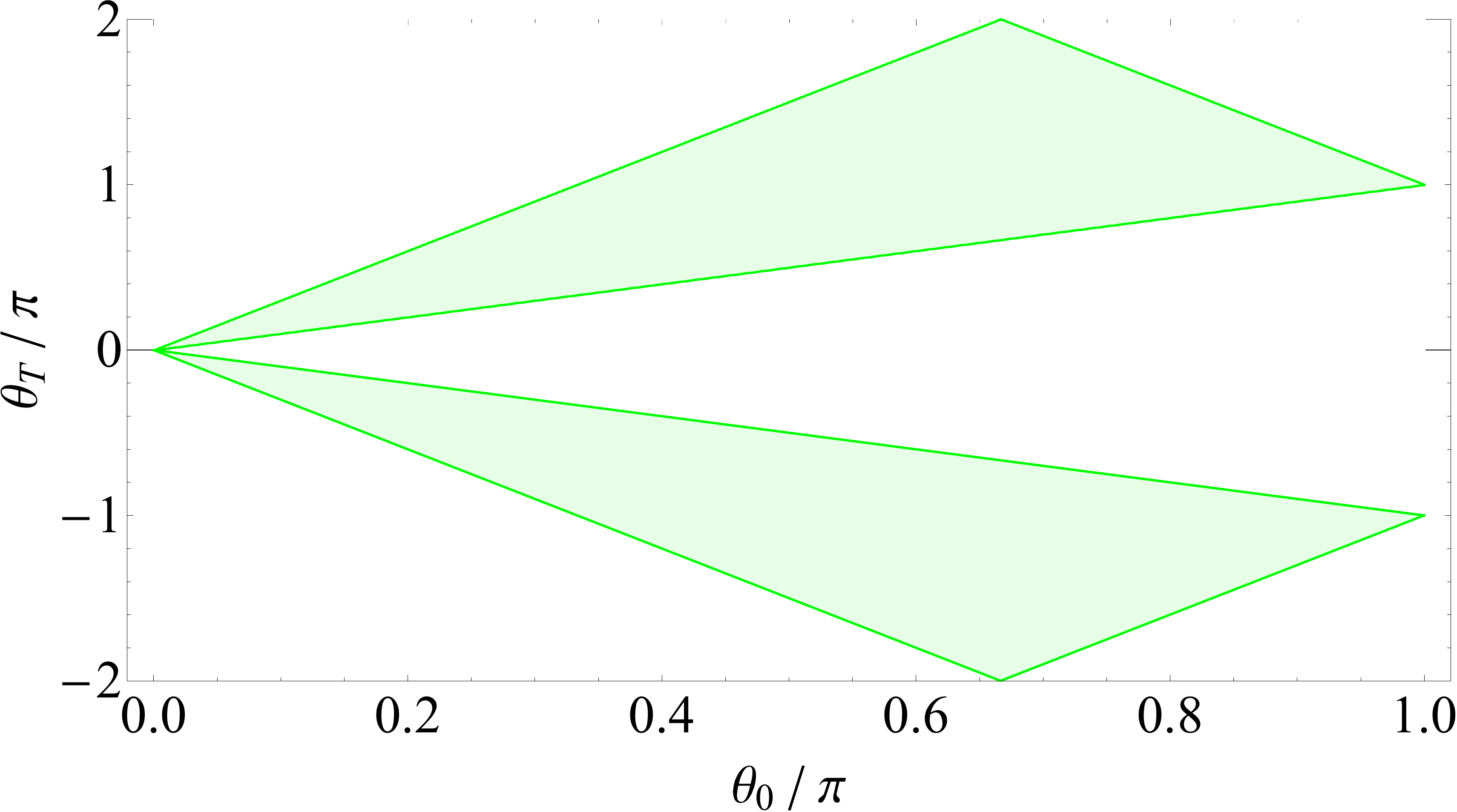}
\caption{\label{Fig:Sequence3}Shaded region indicates $(\theta_0,\theta_T)$ such that the trigonometric polynomial in Eq.~\ref{Eq.LinearSystem3} is achievable by some choice of $\vec{\phi}\in\mathbb{R}^3$.}
\end{figure}

\section{Length $4$ Composite Gates}
When $L=4$, $A[\theta], C[\theta]$ are Fourier series constrained by 
\begin{align}
\label{Eq.LinearSystem4}
1&=a_0+a_1+a_2,
\\ \nonumber
0&=a_0+a_1 \cos{(\theta_0)}+a_2\cos{(2\theta_0)}-\cos{(\theta_T/2)}, 
\\ \nonumber
0&=c_1 \sin{(\theta_0)}+c_2\sin{(2\theta_0)}+\sin{(\theta_T/2)}, 
\\ \nonumber
0&=\cos{\left(\frac{\theta_T}{2}\right)}\left(a_1\sin{(\theta_0)}+2 a_2\sin{(2\theta_0)}\right)+\sin{\left(\frac{\theta_T}{2}\right)}\left(c_1\cos{(\theta_0)}+2 c_2\cos{(2\theta_0)}\right).
\end{align}
Thus we have $4$ linear equations for $5$ free parameters. All that remains is to choose one more linear equation, say $c_2=0$, that leads to the satisfaction of condition Thm.~\ref{Thm:Characterization}.2. Note that many such choices of this last linear equation are possible. We explore two possibilities in the following.

\subsection{Symmetric Length 4 Composite Gate}
One particularly simple choice for the last linear equation is 
\begin{align}
\label{Eq:SymConstraint}
A[\pi]=1 \Rightarrow a_0-a_1+a_2=1.
\end{align}
Together with Eq.~\ref{Eq.LinearSystem4}, solving for $\{a_{0,1,2},c_{1,2}\}$ produces the Fourier series
\begin{align}
\label{Eq:SymPolys}
%
A[\theta]&=1+\frac{\sin^2{\left(\frac{\theta_T}{4}\right)}}{\sin^2{(\theta_0)}}(\cos{(2\theta)}-1)
 \\ \nonumber
%
C[\theta]&=\frac{\tan{\left(\frac{\theta_T}{4}\right)}}{\sin{(\theta_0)}}\left[-2\sin{(\theta)}+\frac{\sin^2{\left(\frac{\theta_T}{4}\right)}}{\sin^2{(\theta_0)}}\left[2\sin{(\theta)}-\cos{(\theta_0)}\sin{(2\theta)}\right]\right].
\end{align}

The phases $\vec{\phi}$ implementing this gate can be efficiently computed in principle from Eq.~\ref{Eq:SymPolys} using Thm.~\ref{Thm:Characterization}, and the parameter space of achievable $(\theta_0,\theta_T)$ can be obtained numerically by checking that the maximum of $A^2[\theta]+C^2[\theta]$ is $1$. 

Alternatively, a more elegant analytic approach is enabled by our use of the necessary and sufficient conditions of Thm.~\ref{Thm:Characterization}. By construction, conditions (1,3,4) of Thm.~\ref{Thm:Characterization} are satisfied by our choice of $A,C$. Thus all that remains to guarantee that $\vec{\phi}$ exists is to check condition (2) that $A^2[\theta]+C^2[\theta]\le 1$. We now derive necessary and sufficient conditions for $(\theta_0,\theta_T)$ that satisfy this condition. Let us expand $A^2[\theta]+C^2[\theta]$ about $\theta=0,\pi$
\begin{align}
A^2[\theta]+C^2[\theta]&=1+\frac{\theta^2}{8}\left[
\left(\cos{(2\theta_0)-\cos{(\theta_T/2)}}\right)\frac{\sin^2{(\theta_T/4)}\tan^2{(\theta_T/4)}}{\sin^2{(\theta_0/2)}\cos^6{(\theta_0/2)}}
\right]+\mathcal{O}(\theta^4)\le 1 \\ \nonumber
A^2[\theta]+C^2[\theta]&=1+\frac{(\theta-\pi)^2}{8}\left[
\left(\cos{(2\theta_0)-\cos{(\theta_T/2)}}\right)\frac{\sin^2{(\theta_T/4)}\tan^2{(\theta_T/4)}}{\sin^2{(\theta_0/2)}\cos^6{(\theta_0/2)}}
\right]+\mathcal{O}((\theta-\pi)^4)\le 1 \\ \nonumber.
\end{align}
By construction through the system of linear constraints, $A^2[\theta]+C^2[\theta]=1$ at $\theta=0,\theta_0,1,2\pi-\theta_0$ and these are also stationary points. By choosing 
\begin{align}
g_4(\theta_0,\theta_T)=\cos{(2\theta_0)-\cos{(\theta_T/2)}} < 0,
\end{align}
the intermediate value theorem tells us that at least $4$ additional stationary points also develop at $A^2[\theta]+C^2[\theta]<1$ between these values of $\theta$. Thus we have identified at least $8$ stationary points of $A^2[\theta]+C^2[\theta]$ that all have value  $\le 1$. However, as $A^2[\theta]+C^2[\theta]$ is a Fourier series of degree $4$, it has at most $8$ stationary points in $\theta\in[0,2\pi)$. Thus we conclude that $\forall\theta\in \mathbb{R},\;A^2[\theta]+C^2[\theta]\le 1$ if $\{(\theta_a,\theta_T)|g_4(\theta_0,\theta_T) < 0\}$. This can be strengthened to if and only if $\{(\theta_a,\theta_T)|g_4(\theta_0,\theta_T) \le 0\}$, plotted in Fig.~\ref{Fig:SymmetricSequence}, by noting that $A^2[\theta]+C^2[\theta]=1$ when $g_4(\theta_0,\theta_T)=0$.
\begin{figure}
\includegraphics[width=0.49\textwidth]{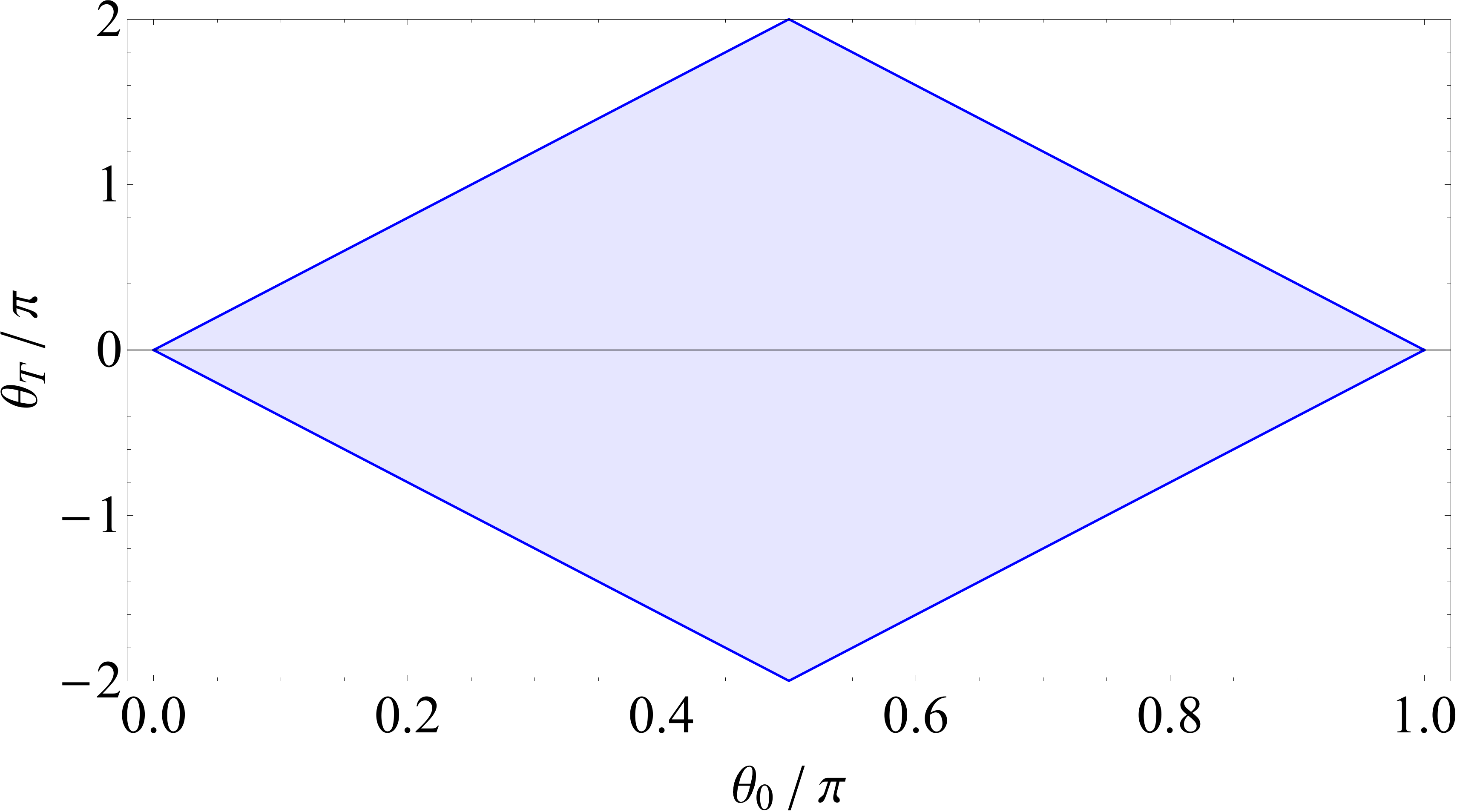}
\includegraphics[width=0.5\textwidth]{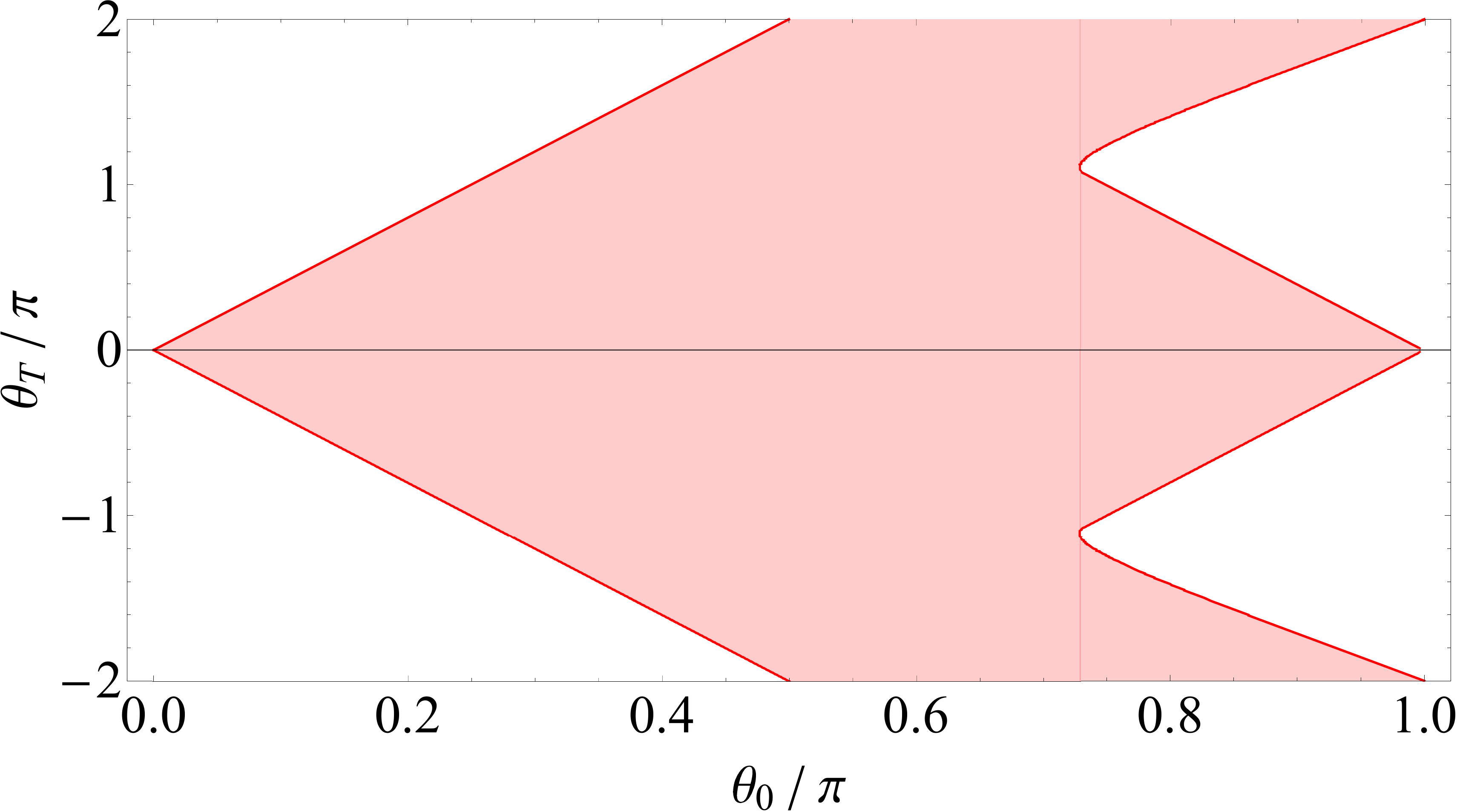}
\caption{\label{Fig:SymmetricSequence}Shaded region indicates $(\theta_0,\theta_T)$ such that the Fourier series in Eq.~\ref{Eq:SymPolys} (left) or Eq.~\ref{Eq:AntiSymmetricConstraint} (right) is achievable by some choice of $\vec{\phi}\in\mathbb{R}^4$.}
\end{figure}

This example also has a simple closed-form solution valid for $\{(\theta_a,\theta_T)|g_4(\theta_0,\theta_T) \le 0 \land \theta_T\ge 0\}$:
\begin{align}
\phi_0&=-\pi/2+\gamma,\quad \phi_1=\pi/2+\gamma+\chi,\quad\phi_2=\phi_1,\quad\phi_3=\phi_0,\\ \nonumber
\chi&=-\cos^{-1}\left[1-2\frac{\sin^2{\left(\frac{\theta_T}{4}\right)}}{\sin^2{(\theta_0)}}\right],\quad 
\gamma=\tan^{-1}\left[\frac{\cos{(\chi/2)}}{\sin{(\chi/2)}}\frac{\cos{(\theta_0)}}{\frac{\sin^2{(\theta_0)}}{\sin^2{\left(\frac{\theta_T}{4}\right)}}-1}\right].
\end{align}
Note that it suffices to consider only this range of $\theta_T\ge 0$ using the identity
\begin{align}
\label{Eq:Identity}
\hat{R}_0[\theta]=\hat{R}_\pi[-\theta]=\hat{R}_\pi[4\pi-\theta].
\end{align}
Thus the phases $\vec{\phi}$ implementing the composite gates for $\theta_T\in[-2\pi,0)$ can be obtained by adding $\pi$ to the phases of composite gates for $\theta_T\in(0,2\pi]$.

Observe from Fig.~\ref{Fig:SymmetricSequence} that the full range of $\theta_T\in[-2\pi,2\pi]$ can only be achieved by symmetric length $4$ composite gates with a base rotation angle of $\theta_0=\pi/2$. In comparison, there exist sequences that implement $\theta_T\in[-\pi,\pi]$ for any $\theta_0\in[\pi/4,3\pi/4]$. Thus by applying this composite gate twice, the full range of $\theta_T$ can be covered by some range of $\theta_0$. 

\begin{figure}[b]
\includegraphics[width=0.48\textwidth]{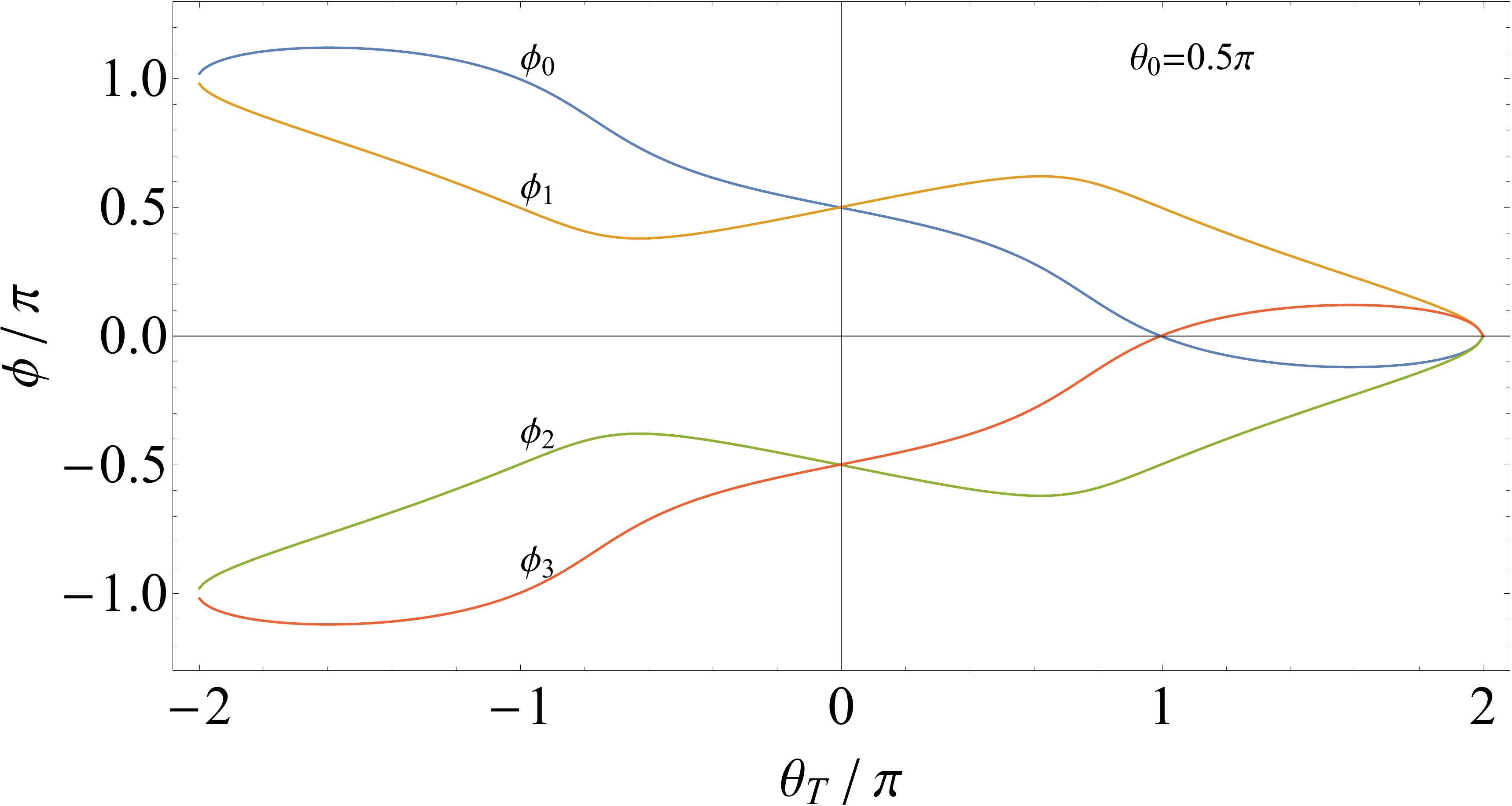}
\includegraphics[width=0.48\textwidth]{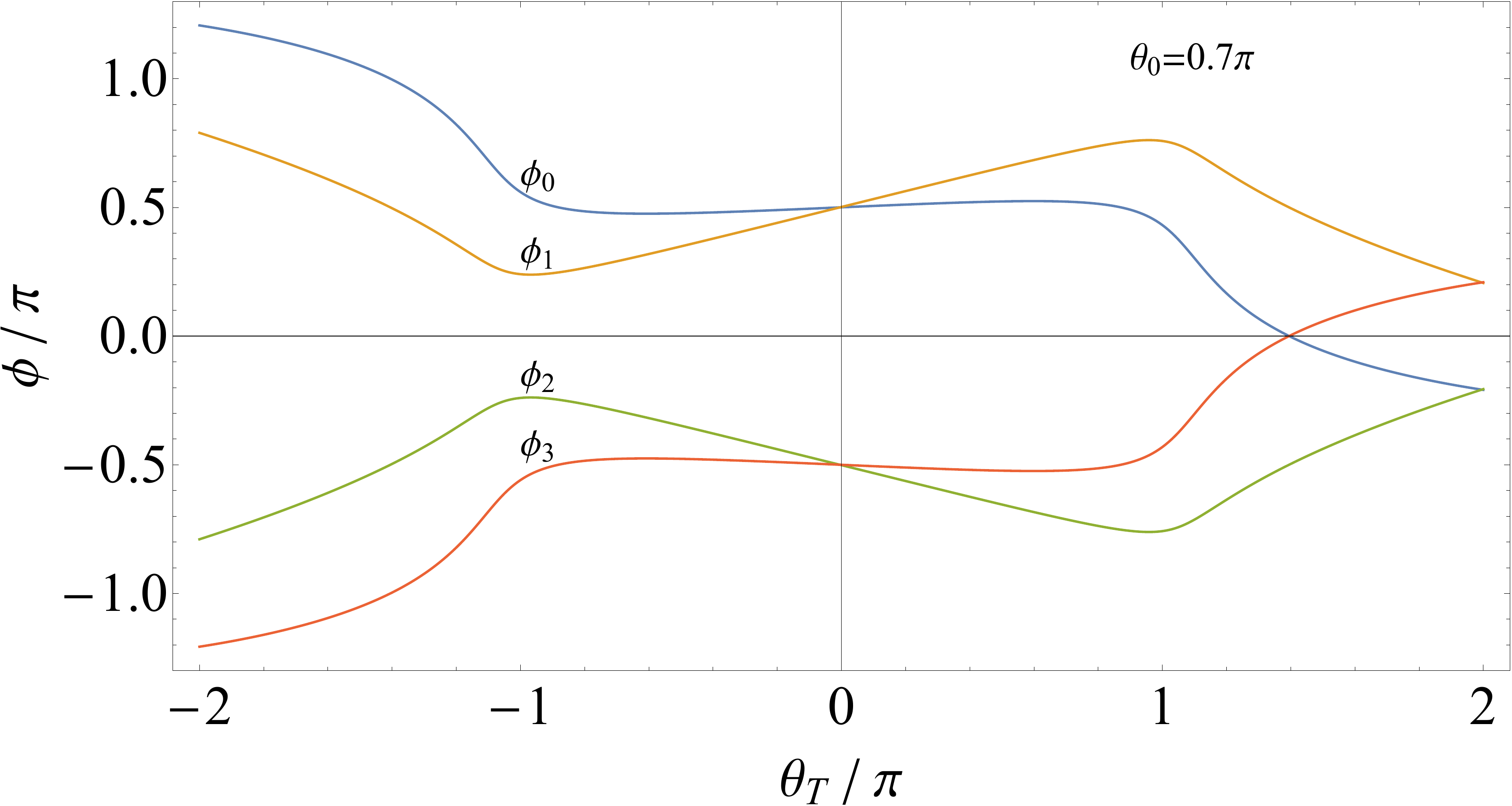}
\caption{\label{Fig:AntiSymmetricSequencePhases}Phases $\vec{\phi}$ implementing the composite gate corresponding to Eq.~\ref{Eq:AntiSymmetricConstraint} for $\theta_0=0.5\pi$ (left) and $\theta_0=0.7\pi$ (right).}
\end{figure}

\subsection{Anti-Symmetric Length 4 Composite Gate}
Another choice for the last linear equation for $0<\theta_T\le 4 \theta_0$ is
\begin{align}
\label{Eq:AntiSymmetricConstraint}
2c_1+4c_2=\cot{\left(\frac{\theta_0}{2}\right)}\tan{\left(\frac{\theta_T}{2}\right)}\left(1-\text{sign}[\sin{(\theta_T/2)}]\sqrt{1+\frac{\cos{(\theta_T/2)}}{\cos^2{(\theta_0/2)}}\left(\frac{1}{\cos^2{(\theta_0/2)}}-\frac{2}{\cos^2{(\theta_T/4)}}\right)}\right).
\end{align}
As described in Eq.~\ref{Eq:Identity}, it suffices to consider only composite gates implementing $\theta_T\in[0,2\pi]$.

Eq.~\ref{Eq:AntiSymmetricConstraint} combined with Eq.~\ref{Eq.LinearSystem4} can in principle be solved for the $\{a_{0,1,2},c_{1,2}\}$ to produce the Fourier series of $A[\theta], C[\theta]$--though an extremely complicated expression, it is nevertheless obtainable in closed form. Thm.~\ref{Thm:Characterization} assures us that for choices of $(\theta_0,\theta_T)$ that satisfy its conditions, $\vec{\phi}$ implementing the composite gate corresponding to Eq.~\ref{Eq:AntiSymmetricConstraint} can be computed using the techniques described in~\cite{Low2016}. 

Though more complicated than the symmetric composite gates, the utility of composite gates corresponding to Eq.~\ref{Eq:AntiSymmetricConstraint} is evident by plotting in Fig.~\ref{Fig:SymmetricSequence} the region of achievable $(\theta_0,\theta_T)$ where $A^2[\theta]+C^2[\theta]\le 1$. Unlike the symmetric composite gates, observe that for a range of base rotation angles $\theta_0\in[\pi/2,0.728\pi]$, the full range of $\theta_T\in[-2\pi,2\pi]$ is accessible.

While $\vec{\phi}$ can in principle be expressed in closed-form as a function of $(\theta_0,\theta_T)$, the result is extremely lengthy and neither insightful nor practical. A more practical option is to compute $\vec{\phi}$, also via~\cite{Low2016}, from $A[\theta], C[\theta]$ with $(\theta_0,\theta_T)$ substituted with numerical values. In this manner, we plot in Fig.~\ref{Fig:AntiSymmetricSequencePhases} $\vec{\phi}$ as a function of $\theta_T$ for $\theta_0=0.5\pi$ and $\theta_0=0.7\pi$. Note that for these composite gates,
\begin{align}
\phi_2=-\phi_1,\quad \phi_3=-\phi_0.
\end{align}

\end{widetext}

\end{document}